\documentclass[12pt]{article}

\usepackage{amsmath}
\usepackage{graphicx}
\usepackage{cite}

\newcommand{\ket}[1]{| #1 \rangle}
\newcommand{\bra}[1]{\langle #1 |}
\newcommand{\hcs}[1]{#1^\dagger #1}
\newcommand{\expv}[1]{\langle #1 \rangle}

\begin{document}

\begin{center}
{\Large\bf Optimal quantum measurements
for additive information and disturbance measures}
\vskip .6 cm
Hiroaki Terashima
\vskip .4 cm
{\it Department of Science Education, Cooperative Faculty of Education, \\
Gunma University, \\
Maebashi, Gunma 371-8510, Japan}
\vskip .6 cm
\end{center}

\begin{abstract}
Additive measures for
information and disturbance in quantum measurements of a system
are defined from well-known multiplicative measures
such as estimation and operation fidelities using a logarithm.
This is motivated by the fact that
information and disturbance are naturally assumed to be additive
while performing independent measurements on separable systems.
Although the additivity makes no remarkable difference
when information and disturbance are separately considered,
it can change measurements
that only introduce minimal disturbance relative to
the amount of information.
Such optimal measurements are shown
for additive information and disturbance measures
with a tradeoff relationship.
\end{abstract}

\section{Introduction}
A quantum measurement that provides information about a system
inevitably introduces disturbance to the state of the system.
Although, in general, measurements are desired to be
highly informative and minimally disturbing,
a tradeoff exists such that
the more information a measurement provides,
the greater the disturbance.
This tradeoff relationship is
saturated by optimal measurements---those that provide
a specific amount of information
with minimum disturbance.
Optimal measurements are of great interest
in quantum measurement theory
and in the realization of
quantum information processing and communication.
The optimal measurements depend on
the way information and disturbance are quantified.
Various information and disturbance measures
have been used to discuss
the diverse tradeoffs between information and
disturbances~\cite{FucPer96,Banasz01,FucJac01,BanDev01,%
DArian03,Ozawa04,GenPar05,MiFiFi05,Maccon06,Sacchi06,BusSac06,Banasz06,%
BuHaHo07,CheLee12,RenFan14,FGNZ15,ShKuUe16,%
ShaPat18,LiuDeB21,LeKiNh21,SuLiLu22,SunLi24,SaAkSh24}.

For example,
estimation fidelity~\cite{Banasz01},
operation fidelity~\cite{Banasz01},
and physical reversibility~\cite{KoaUed99}
are a few well-known measures in quantum measurements.
The estimation fidelity $G$ is an information measure,
which is determined by the overlap between the original and
estimated states from the measurement outcome.
By contrast,
the operation fidelity $F$ is a disturbance measure,
which is defined by the overlap between the original and
disturbance states.
The physical reversibility $R$ is also considered
a disturbance measure;
it represents the maximum probability
of recovering the original state from the disturbed state
using a reversing measurement~\cite{UeImNa96,Ueda97}.
Tradeoffs have been shown
between $G$ and $F$~\cite{Banasz01} and
between $G$ and $R$~\cite{CheLee12},
and have been verified~\cite{SRDFM06,BaChKi08,CZXTLX14,LRHLK14}
using single-photon experiments.
Recently, a triplewise tradeoff among $G$, $F$, and $R$
has been shown~\cite{LeKiNh21}
and experimentally verified in Ref.~\cite{HKCKLL22}.

However, these measures are not additive.
Consider performing two independent measurements
on two separable systems.
The total information gain is naturally assumed to be
the addition of the two individual information gains.
However, this additivity is not valid
for the estimation fidelity.
The total estimation fidelity is
not the sum but the product
of the two individual estimation fidelities.
In other words, the estimation fidelity is multiplicative;
therefore, it cannot be regarded
as a natural information measure.
However, its logarithm is additive,
thus making it a candidate for natural information measure.
Moreover, disturbance is naturally assumed to be additive
if it is in a tradeoff relationship with information,
whereas operation fidelity
is not additive but multiplicative.

Herein, we define
the additive information and disturbance measures
and determine the optimal measurements.
Three additive measures $I_G$, $D_F$, and $D_R$
are defined by the logarithms of $G$, $F$, and $R$, respectively.
Because the logarithm is a monotonically increasing function,
the additivity does not make any remarkable difference
when information and disturbance are separately considered.
However, it affects the measurements
when information and disturbance are considered together.
By incorporating a reduction of
the Shannon entropy~\cite{DArian03,Terash15}, $I$,
as an inherently additive information measure,
four additive information--disturbance pairs are considered:
$I_G$--$D_F$, $I_G$--$D_R$, $I$--$D_F$, and $I$--$D_R$.

For each information--disturbance pair,
optimal measurements are determined using
a physically allowed region~\cite{Terash16}
on the information--disturbance plane.
This region is obtained by
plotting all physically possible
single-outcome measurement processes on the plane.
The curvature of the boundary determines
the optimal measurements for the information--disturbance pair.
While the optimal measurements for $I_G$--$D_F$
are the same as those for $G$--$F$,
the optimal measurements for $I_G$--$D_R$
are the same as those for $G$--$F$ not for $G$--$R$.
The optimal measurements for $G$--$R$
are known to be not necessarily
optimal for $G$--$F$~\cite{CheLee12,LRHLK14};
however, the optimal measurements for $I_G$--$D_R$ are always
optimal for $I_G$--$D_F$.
In addition,
the optimal measurements for $I$--$D_F$ have
the same form as those for $I$--$F$;
however, the optimal measurements for $I$--$D_R$
are completely different from those for $I$--$R$.

The remainder of this paper is organized as follows.
Section~\ref{sec:preliminary} reviews
the four well-known information and disturbance measures.
Section~\ref{sec:additive} introduces two each additive information
and disturbance measures.
Section~\ref{sec:region} elucidates the physically allowed regions
of the additive information and disturbance.
Section~\ref{sec:optimal} determines the optimal measurements
for the additive information and disturbance.
Finally, Section~\ref{sec:summary} presents the summary.

\section{\label{sec:preliminary}Preliminaries}
First, we recall some extensively used
information and disturbance measures
in the quantum measurement theory.
An ideal quantum measurement~\cite{NieCav97} is described by
a set of measurement operators $\{\hat{M}_m\}$~\cite{NieChu00}
that satisfies
\begin{equation}
\sum_m\hcs{\hat{M}_m}=\hat{I},
\label{eq:complete}
\end{equation}
where $\hat{I}$ denotes the identity operator.
The measurement operator $\hat{M}_m$ describes
the measurement process of outcome $m$.
If the system is in a state $\ket{\psi(a)}$,
the measurement on the system yields the outcome $m$
with the probability
\begin{equation}
p(m|a)=\bra{\psi(a)}\hcs{\hat{M}_m}\ket{\psi(a)},
\end{equation}
changing the state of the system to
\begin{equation}
\ket{\psi(m,a)}=\frac{1}{\sqrt{p(m|a)}}\,\hat{M}_m\ket{\psi(a)}.
\end{equation}
When the system is initially assumed to be in a state $\ket{\psi(a)}$
with probability $p(a)$,
the total probability of $m$ is given by
\begin{equation}
p(m)=\sum_{a} p(m|a)\,p(a).
\end{equation}

The amount of information provided by the measurement
can be quantified using two measures:
estimation fidelity~\cite{Banasz01} and
entropy reduction~\cite{DArian03,Terash15}.
The estimation fidelity defines
the similarity between the premeasurement and best-estimated states
from the outcome.
It is given by
\begin{equation}
 G =\sum_{m,a} p(a)\,p(m|a)\,\bigl|\expv{\psi(a_m)|\psi(a)}\bigr|^2,
\label{eq:defG}
\end{equation}
where $a_m$ is $a$ that maximizes $p(m|a)$.
The entropy reduction defines
the amount of the outcome to be compensated for
lack of information about the state of the system.
The outcome $m$ brings some information to change
the state probability distribution from $\{p(a)\}$
to $\{p(a|m)\}$,
where
\begin{equation}
  p(a|m) =\frac{p(m|a)\,p(a)}{p(m)}
\label{eq:defPam}
\end{equation}
is the conditional probability of
premeasurement state $\ket{\psi(a)}$ given outcome $m$.
This change decreases the Shannon entropy by
\begin{equation}
  I  =\sum_{m} p(m)\left[-\sum_a p(a)\log_2 p(a)
         +\sum_{a} p(a|m)\log_2 p(a|m)\right]
\label{eq:defI}
\end{equation}
as the entropy reduction.

By contrast,
the degree of disturbance owing to the measurement
can be quantified using two measures:
operation fidelity~\cite{Banasz01}
and physical reversibility~\cite{KoaUed99}.
The operation fidelity defines
the similarity between the pre- and postmeasurement states.
It is given by
\begin{equation}
 F=\sum_{m,a} p(a)\,p(m|a)\,\bigl|\expv{\psi(m,a)|\psi(a)}\bigr|^2.
\label{eq:defF}
\end{equation}
The physical reversibility defines
how effortlessly the measured system can be restored to
the premeasurement state
using a reversing measurement~\cite{UeImNa96,Ueda97}.
The best reversing measurement
restores the state of the system
with a probability of
\begin{equation}
   R=\sum_m \inf_{\ket{\psi}}\, \bra{\psi}\hcs{\hat{M}_m}\ket{\psi}.
\label{eq:defR}
\end{equation}
This probability is regarded as physical reversibility.

\section{\label{sec:additive}Additive Measures}
Next, we check the additivity of the above measures
and derive the additive measures from the nonadditive ones.
To check the additivity,
two independent measurements were performed on
two separable systems.
These two measurements can be regarded
as a composite measurement on a composite system
using the tensor products of states and operators.

For example,
the estimation fidelity $G$ is elucidated as follows.
Let $G_1$ and $G_2$ be
the estimation fidelities of the two individual measurements.
By employing the tensor products,
the estimation fidelity of the composite measurement
is shown to be $G_\mathrm{c}=G_1G_2$.
In general, $G$ is not additive but multiplicative;
thus, it is unnatural as an information measure.
When the net amount of information is extracted,
$G$ worsens.
The identity operation, as a kind of measurement,
provides no information
but gives $G=1/d$ with $d$ being
the dimension of the Hilbert space of the system.
Therefore,
the net amount of information should be $G-(1/d)$.
Moreover, this is not multiplicative.

An additive measure can be defined from $G$
using the logarithm.
A naive candidate is $\log_2(d G)$;
it is positive for any measurement and
zero for the identity operation
because of $d$ in the argument.
However,
$\log_2(d G)$ is still unnatural
because it is not additive among outcomes.
Let $G(m)$ be the estimation fidelity
when a single outcome $m$ is obtained from the measurement.
As $G=\sum_m p(m)\,G(m)$,
$\log_2(d G)$ cannot 
be decomposed into the amount of information for each outcome.

Therefore,
an additive measure must be defined
from $G(m)$ rather than $G$.
Using~\cite{Terash15}
\begin{equation}
 G(m)=\sum_a p(a|m)\,\bigl|\expv{\psi(a_m)|\psi(a)}\bigr|^2,
\label{eq:defGm}
\end{equation}
we can define the additive measure of
the information provided by a single outcome $m$
as
\begin{equation}
 I_G(m) \equiv\log_2 [d G(m)].
\label{eq:defIGm}
\end{equation}
The average over the outcomes is
\begin{equation}
 I_G \equiv\sum_m p(m)\,\log_2 [d G(m)].
\label{eq:defIG}
\end{equation}
This measure satisfies
\begin{equation}
   0 \le I_G(m) \le \log_2 \left(\frac{2d}{d+1}\right),
\end{equation}
where the minimum is achieved by the identity operation and
the maximum by the rank-$1$ projective measurement.

By contrast,
the entropy reduction $I$ is additive
because it originally contains the logarithm.
Thus, another additive information measure
related to a single outcome is~\cite{DArian03,Terash15}
\begin{equation}
 I(m) =-\sum_a p(a)\log_2 p(a)
         +\sum_{a} p(a|m)\log_2 p(a|m),
\label{eq:defIm} 
\end{equation}
and its average over outcomes is $I$.
This measure satisfies~\cite{Terash15}
\begin{equation}
 0 \le I(m) \le\log_2d-\frac{1}{\ln2}
        \left(\frac{1}{2}+\frac{1}{3}+\cdots+\frac{1}{d}\right).
\end{equation}

Similarly, the operation fidelity $F$ and
physical reversibility $R$ can be elucidated.
The net degrees of disturbance, $1-F$ and $1-R$,
are neither additive nor multiplicative.
Notably, their minus signs are because
$F$ and $R$ decrease as the disturbance increases.
Moreover, the logarithms $-\log_2 F$ and $-\log_2 R$
are not additive among the outcomes.

Thus, the additive disturbance measures should be
defined from $F(m)$ and $R(m)$,
where~\cite{Terash15}
\begin{align}
 F(m) &=\sum_a p(a|m)\,\bigl|\expv{\psi(m,a)|\psi(a)}\bigr|^2,
         \label{eq:defFm} \\
 R(m) &=\sum_a p(a|m)\,\frac{\inf_{\ket{\psi}}\,\bra{\psi}
                              \hcs{\hat{M}_m}\ket{\psi}}{p(m|a)}
         \label{eq:defRm} 
\end{align}
are the operation fidelity and physical reversibility
when a single outcome $m$ is obtained from measurement.
The additive measures of a single outcome are given by
\begin{align}
 D_F(m) &\equiv -\log_2 F(m),
         \label{eq:defDFm} \\
 D_R(m) &\equiv-\log_2 R(m)
         \label{eq:defDRm}
\end{align}
and their averages are
\begin{align}
 D_F &\equiv-\sum_m p(m)\,\log_2 F(m),
         \label{eq:defDF} \\
 D_R &\equiv-\sum_m p(m)\,\log_2 R(m).
         \label{eq:defDR}
\end{align}
These measures satisfy
\begin{align}
   0 &\le D_F(m) \le \log_2 \left(\frac{d+1}{2}\right), \\
   0 &\le D_R(m) < \infty,
\end{align}
where a unitary part of $\hat{M}_m$ irrelevant to
information gain~\cite{Terash15} is removed in $D_F(m)$
to determine the inevitable state change
owing to extracting information.

\section{\label{sec:region}Physically Allowed Regions}
Using additive measures,
we present a physically allowed region
of information and disturbance~\cite{Terash16}
on the information--disturbance plane.
As the two information measures $\{I_G,I\}$ and
two disturbance measures $\{D_F,D_R\}$ have already been introduced,
four information--disturbance pairs can be considered.
They are calculated for
physically possible measurements
on a system in a completely unknown pure state.
The formulas for the original measures are
presented in Ref.~\cite{Terash15}
as the functions of the singular values of $\hat{M}_m$.

\begin{figure}
\begin{center}
\includegraphics[scale=0.52]{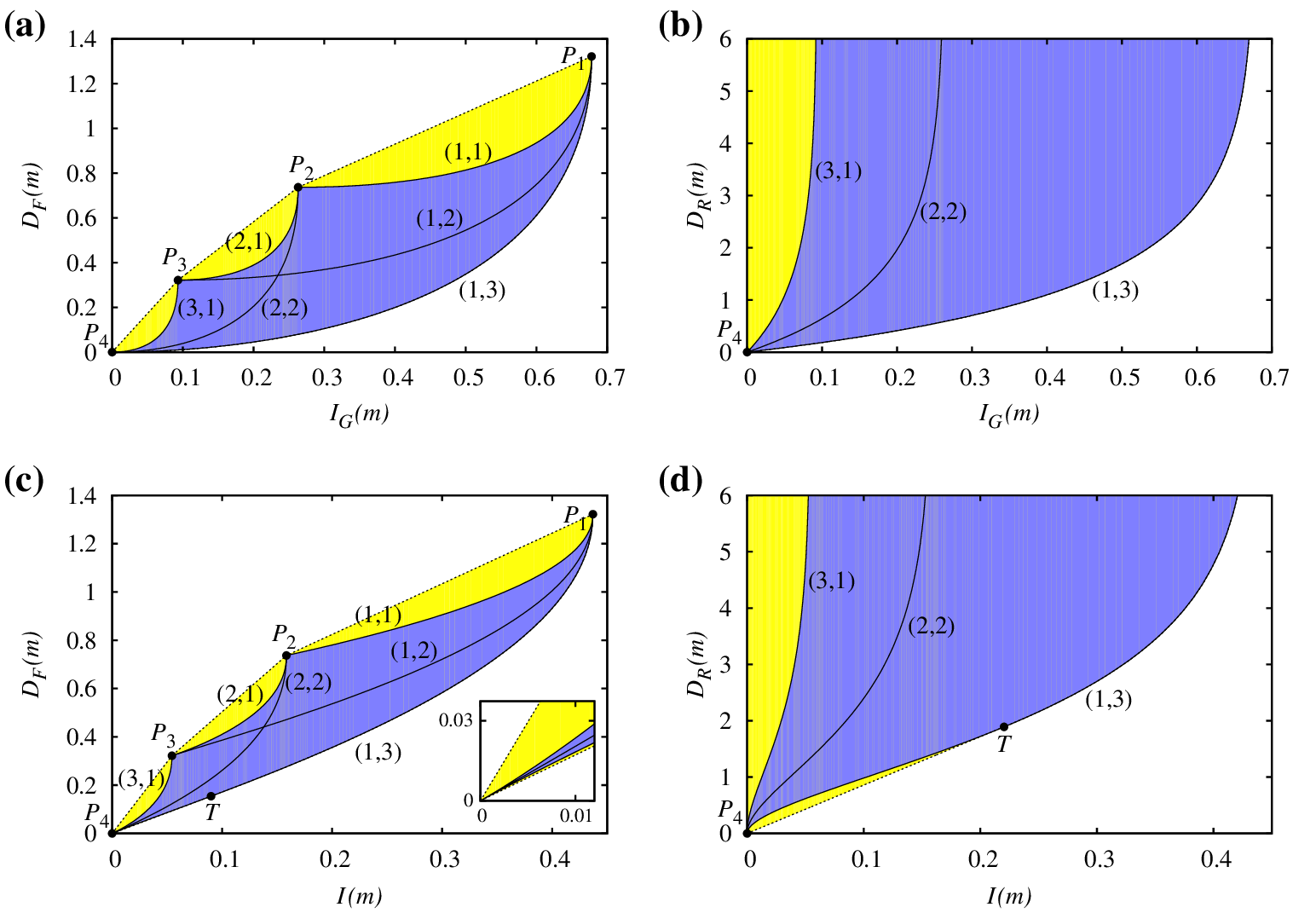}
\end{center}
\caption{\label{fig1}
Physically allowed region of information and disturbance for $d=4$.
Two information measures $\{I_G(m),I(m)\}$
and two disturbance measures $\{D_F(m),D_R(m)\}$ are used in four pairs:
(a) $I_G(m)$--$D_F(m)$,
(b) $I_G(m)$--$D_R(m)$,
(c) $I(m)$--$D_F(m)$, and
(d) $I(m)$--$D_R(m)$.
The blue (dark gray) and yellow (light gray) shaded parts represent
the region of a single outcome and
region extended by averaging over the outcomes, respectively.
The point $P_r$ corresponds to
the measurement operator $\hat{P}^{(d)}_{r}$, and
the line $(k,l)$ corresponds to $\hat{M}^{(d)}_{k,l}(\lambda)$.
The points $T$ in panels (c) and (d)
denote the point of tangency of
the tangent line drawn from $P_d$ to $(1,d-1)$.
}
\end{figure}%
If a single outcome $m$ is to be obtained from measurement,
the allowed region can easily be shown
by plotting all physically possible $\hat{M}_m$s
on the information--disturbance plane~\cite{Terash16}.
Figure~\ref{fig1} depicts the results when $d=4$ in blue (dark gray)
for the four pairs:
$I_G(m)$--$D_F(m)$,
$I_G(m)$--$D_R(m)$,
$I(m)$--$D_F(m)$, and
$I(m)$--$D_R(m)$.
In this figure,
$P_r$ denotes the point corresponding
to $\hat{M}_m=c\hat{P}^{(d)}_{r}$,
where $c$ is a constant and
$\hat{P}^{(d)}_{r}$ is a projective operator of rank $r$.
In an orthonormal basis $\{\ket{i}\}$ ($i=1,2,\ldots,d$)
of the Hilbert space of the system,
the projective operator can be written as
\begin{equation}
\hat{P}^{(d)}_{r}=\sum_{i=1}^{r} \ket{i}\bra{i}.
\end{equation}
The factor $c$ comes from the rescaling invariance
of the original measures~\cite{Terash15}.
Similarly, $(k,l)$ denotes the line corresponding to
$\hat{M}_m=c\hat{M}^{(d)}_{k,l}(\lambda)$ with $0\le\lambda\le1$,
where $\hat{M}^{(d)}_{k,l}(\lambda)$ can be written as
\begin{equation}
\hat{M}^{(d)}_{k,l}(\lambda)=\sum_{i=1}^{k} \ket{i}\bra{i}
                       +\sum_{i=k+1}^{k+l} \lambda \ket{i}\bra{i}.
\label{eq:Mdkl}
\end{equation}
The lower boundary of the region is $(1,d-1)$
between $P_d$ and $P_1$ in all panels.
The upper boundary comprises $(k,1)$
for $k=d-1,d-2,\ldots,1$ in panels (a) and (c),
whereas it is $(d-1,1)$ between $P_d$ and $P_{d-1}$
and $D_R(m)=\infty$ between $P_{d-1}$ and $P_1$
in panels (b) and (d).
Notably, $P_r$ for $r\neq d$
and $(k,l)$ for $k+l \neq d$ cannot be shown
in panels (b) and (d) because $D_R(m)=\infty$.
The regions are similar to those
of the original measures~\cite{Terash16}
when vertically flipped.

When the information and disturbance
are averaged over the outcomes,
the physically allowed region can be derived from
the region of a single outcome
using an analogy with the center of mass
of particles~\cite{Terash16}.
In general, a measurement is represented by
a set of measurement operators $\{\hat{M}_m\}$,
each of which corresponds to a point $R_m$
in the region of a single outcome
with weight $p(m)$.
This is analogous to a set of particles,
each of which is at the point $R_m$ with mass $p(m)$.
The center of mass of these particles indicates
the average information and disturbance of $\{\hat{M}_m\}$.
Therefore, the region for the average values
is given by the convex hull of
the region of a single outcome.
In other words, averaging over outcomes extends the region
by replacing the concave parts of the boundary
with straight lines,
as shown in yellow (light gray) in Fig.~\ref{fig1}.

As shown in Fig.~\ref{fig1}(c) and (d),
the region is extended a little below
because $(1,d-1)$ has a slight dent near $P_d$
to be inverted S-shaped.
In fact, from the derivatives of the original measures~\cite{Terash18},
the second derivative of $(1,d-1)$ near $P_d$
is calculated to be
\begin{equation}
 \frac{\mathrm{d}^2D_F(m)}{\mathrm{d}I(m)^2}<0
\end{equation}
for $d\ge3$ in $I(m)$--$D_F(m)$ and
\begin{equation}
 \frac{\mathrm{d}^2D_R(m)}{\mathrm{d}I(m)^2}<0
\end{equation}
for any $d$ in $I(m)$--$D_R(m)$.
Therefore, the lower boundary is replaced with
the tangent line drawn from $P_d$ to $(1,d-1)$
between $P_d$ and the point of tangency $T$.
The points $T$ correspond to the measurement operators
$c\hat{M}^{(4)}_{1,3}(0.470)$
and $c\hat{M}^{(4)}_{1,3}(0.291)$
in Fig.~\ref{fig1}(c) and (d),
respectively.

\begin{figure}
\begin{center}
\includegraphics[scale=0.6]{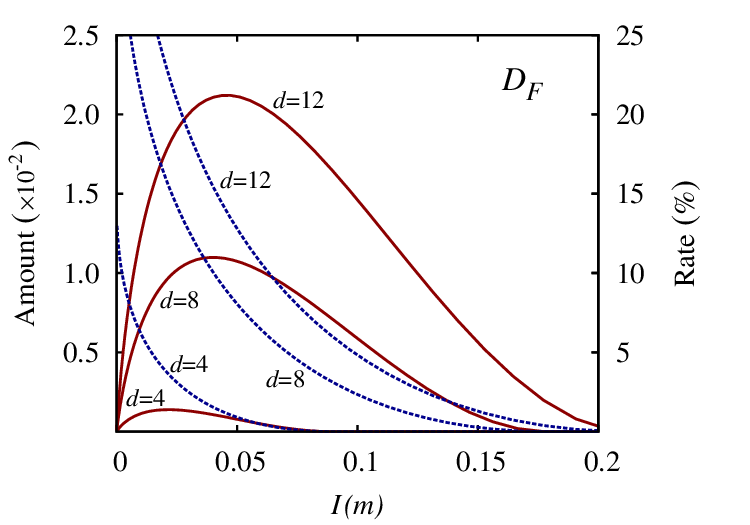}
\end{center}
\caption{\label{fig2}
Amount and rate of decrease in $D_F$
from $(1,d-1)$ to the tangent line between $P_d$ and $T$
when $d=4,8,12$.
The solid and dashed lines denote the amount and rate, respectively.
}
\end{figure}%
\begin{figure}
\begin{center}
\includegraphics[scale=0.6]{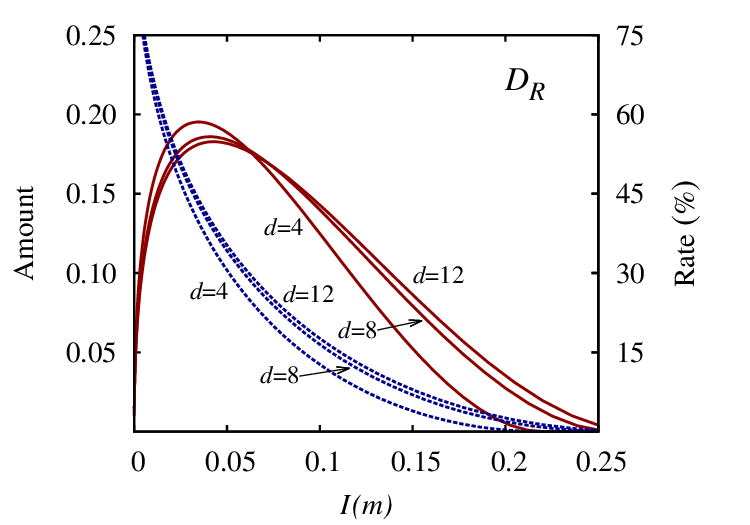}
\end{center}
\caption{\label{fig3}
Amount and rate of decrease in $D_R$
from $(1,d-1)$ to the tangent line between $P_d$ and $T$
when $d=4,8,12$.
The solid and dashed lines denote the amount and rate, respectively.
}
\end{figure}%
Figures~\ref{fig2} and \ref{fig3} show
the decrease in $D_F$ and $D_R$
from $(1,d-1)$ to the tangent line between $P_d$ and $T$
when $d=4,8,12$.
The solid and dashed lines denote the decrease amount and rate,
respectively.
Both the decrease rates increase
as $I(m)$ decreases or $d$ increases
although the rate of $D_R$ is much larger than that of $D_F$.

The lower boundary for the average values
indicates a tradeoff relationship between information and disturbance.
The lower boundary for $I_G$--$D_F$
is given by $(1,d-1)$ as that for $G$--$F$.
Therefore, the tradeoff between $I_G$ and $D_F$
is derived from that between $G$ and $F$~\cite{Banasz01}
by replacing $G$ with $2^{I_G}/d$
and $F$ with $1/2^{D_F}$.
In other words, $I_G$ and $D_F$ satisfy the inequality
\begin{equation}
\sqrt{\frac{1}{2^{D_F}}-\frac{1}{d+1}}\le
\sqrt{\frac{2^{I_G}}{d}-\frac{1}{d+1}}+
\sqrt{(d-1)\left(\frac{2}{d+1}-\frac{2^{I_G}}{d}\right)}.
\label{eq:ineqGF}
\end{equation}
Similarly,
the tradeoff between $I_G$ and $D_R$
is derived from that between $G$ and $R$~\cite{CheLee12}
by replacing $G$ with $2^{I_G}/d$
and $R$ with $1/2^{D_R}$.
Thus, $I_G$ and $D_R$ satisfy the inequality
\begin{equation}
 (d+1)2^{I_G}+(d-1)\frac{1}{2^{D_R}}\le 2d.
\label{eq:ineqGR}
\end{equation}
However, the tradeoff between $I$ and $D_F$
cannot be expressed by a single explicit inequality
because $I$ for $(1,d-1)$ is
a highly complicated function of $\lambda$~\cite{Terash16}
and the lower boundary between $P_d$ and $T$ is
a straight line not $(1,d-1)$.
Let $\mathcal{D}_F(I)$ be a function
from $I$ to $D_F$ on $(1,d-1)$
and $I_T$ be $I$ at $T$.
The tradeoff between $I$ and $D_F$ is then given as
\begin{equation}
 D_F \ge
\begin{cases}
        k_F I            & \mbox{(if $I< I_T$)} \\
        \mathcal{D}_F(I) & \mbox{(if $I\ge I_T$)},
    \end{cases}
\label{eq:ineqIF}
\end{equation}
where $k_F=\mathcal{D}_F(I_T)/I_T$ denotes
a constant of proportionality.
Similarly,
the tradeoff between $I$ and $D_R$ is given as
\begin{equation}
 D_R \ge
\begin{cases}
        k_R I            & \mbox{(if $I< I_T$)} \\
        \mathcal{D}_R(I) & \mbox{(if $I\ge I_T$)}.
    \end{cases}
\label{eq:ineqIR}
\end{equation}

\section{\label{sec:optimal}Optimal Measurements}
Finally, we find the optimal measurements
for the additive information and disturbance measures.
For a specific amount of information,
an optimal measurement is one that
causes the least possible disturbance,
thereby saturating the tradeoff relationship
between information and disturbance.
In other words, the optimal measurements correspond to
the points on the lower boundary of
the physically allowed region for the average values.

The optimal measurements can easily be obtained
using the analogy of
the center of mass of the particles~\cite{Terash16}.
It is represented as
a set of particles
whose center of mass is on the lower boundary.
As shown in Fig.~\ref{fig1}(a) and (b),
the lower boundary for $I_G$--$D_F$ and $I_G$--$D_R$
is just $(1,d-1)$ because $(1,d-1)$ is convex.
For the center of mass to be at a point on a convex boundary,
all particles must be at that point.
Therefore, for an optimal measurement,
all measurement operators should correspond to
an identical point on $(1,d-1)$.
This means that
the optimal measurement $\{\hat{M}_m\}$
comprises the measurement operators
of the form $c\hat{M}^{(d)}_{1,d-1}(\lambda)$
with the same $\lambda$.
For example,
an optimal and minimal measurement
for a given information $I_G$ is
a $d$-outcome measurement
\begin{equation}
\left\{\hat{M}_{1}^{(d)}(\lambda),\hat{M}_{2}^{(d)}(\lambda),
  \ldots, \hat{M}_{d}^{(d)}(\lambda) \right\},
\label{eq:optMm}
\end{equation}
where
\begin{equation}
  \hat{M}_{m}^{(d)}(\lambda)\equiv\frac{1}{\sqrt{1+(d-1)\lambda^2}}
  \left(\ket{m}\bra{m}+\sum_{i\neq m} \lambda \ket{i}\bra{i}\right),
\label{eq:optOp}
\end{equation}
and $\lambda$ is chosen such that
the measurement provides the information $I_G$.
The operator in Eq.~(\ref{eq:optOp}) is proportional
to $\hat{M}^{(d)}_{1,d-1}(\lambda)$,
although the orthonormal basis is rearranged from Eq.~(\ref{eq:Mdkl})
to satisfy Eq.~(\ref{eq:complete}).
This rearrangement does not alter
$I_G(m)$, $D_F(m)$, and $D_R(m)$
from the invariance of the original measures~\cite{Terash15}.
The factor in Eq.~(\ref{eq:optOp}) can also be determined by
Eq.~(\ref{eq:complete}).
This measurement simultaneously saturates
the inequalities in Eqs.~(\ref{eq:ineqGF}) and (\ref{eq:ineqGR})
and the inequality between $G$ and $F$ in Ref.~\cite{Banasz01}.

By contrast,
the lower boundary for $I$--$D_F$ and $I$--$D_R$
is $(1,d-1)$ between $T$ and $P_1$
but a straight line between $P_d$ and $T$
because $(1,d-1)$ is inverted S-shaped,
as shown in Fig.~\ref{fig1}(c) and (d).
Therefore,
depending on the given information $I$,
the optimal measurements can be divided into two cases.
If $I\ge I_T$,
the measurement in Eq.~(\ref{eq:optMm}) is still optimal,
although $\lambda$ is chosen such that
the measurement provides the information $I$.
However, if $I<I_T$,
it causes a little higher disturbance than the minimum
shown by the straight line between $P_d$ and $T$.
Notably, when the center of mass is
at a point on a straight boundary
lying outside the region where the particles can be,
every particle should be located
at either end of the straight section.
Such particles
represent a measurement with measurement operators
corresponding to either $P_d$ or $T$.
In other words, the measurement operators are either
$c\hat{I}$ or
$c\hat{M}^{(d)}_{1,d-1}(\lambda_T)$,
where $\lambda_T$ is $\lambda$ at $T$.
An example of
an optimal and minimal measurement is
a $(d+1)$-outcome measurement
\begin{equation}
\left\{c\hat{M}_{1}^{(d)}(\lambda_T),
  c\hat{M}_{2}^{(d)}(\lambda_T),
  \ldots, c\hat{M}_{d}^{(d)}(\lambda_T),
 \sqrt{1-c^2}\,\hat{I}
 \right\},
\label{eq:optMm2}
\end{equation}
where $c=\sqrt{I/I_T}$.
If $I<I_T$, this measurement
can reduce the disturbance compared with
the measurement in Eq.~(\ref{eq:optMm})
at the rate shown in Figs.~\ref{fig2} and \ref{fig3}.
For example,
when $d=4$ and $I=0.05$,
it decreases $D_F$ and $D_R$ by about 0.8\% and 30\%, respectively.
The inequalities in Eqs.~(\ref{eq:ineqIF}) and (\ref{eq:ineqIR})
are saturated by the measurements in Eq.~(\ref{eq:optMm}) if $I\ge I_T$
and Eq.~(\ref{eq:optMm2}) if $I<I_T$.
However, the two inequalities
are not necessarily saturated at the same time,
as $I_T$ is not equal between $I$--$D_F$ and $I$--$D_R$.

\begin{table}
\caption{\label{tbl1}
Sign of the curvature of $(1,d-1)$.}
\begin{center}
\begin{tabular}{ccccc}
\hline\noalign{\smallskip}
      & \multicolumn{4}{c}{disturbance} \\
\noalign{\smallskip}  \cline{2-5}\noalign{\smallskip}
    information & $1-F(m)$ & $D_F(m)$ & $1-R(m)$ & $D_R(m)$  \\
\noalign{\smallskip}\hline\noalign{\smallskip}
        $G(m)$   & $+$ &  &  $0$  &   \\
        $I_G(m)$  &   &  $+$ &  & $+$ \\
        $I(m)$  & $\mp$ & $\mp$ & $-$ & $\mp$ \\
\noalign{\smallskip}\hline
\end{tabular}
\end{center}
\end{table}%
In general, the optimal measurements are
determined by the sign of the curvature
(i.e., the second derivative) of $(1,d-1)$.
The signs are summarized in Table~\ref{tbl1}
in the form of additive measures and
their original versions~\cite{Terash18}.
The table includes four cases: $+$, $-$, $0$, and $\mp$.
If two information--disturbance pairs have the same sign,
their optimal measurements have the same form.
For example, the optimal measurements coincide
among $G$--$(1-F)$, $I_G$--$D_F$, and $I_G$--$D_R$
because they have the same sign $+$.
If an information--disturbance pair
has the sign $0$ or $-$,
the optimal measurements for that pair coincide with
those for $G$--$(1-R)$ or $I$--$(1-R)$, respectively,
which are also found using the analogy with
the center of mass of particles~\cite{Terash16}.
Although the optimal measurements for
$I$--$(1-F)$, $I$--$D_F$, and $I$--$D_R$ have the same form
because of the same sign $\mp$,
not all coincide because their $I_T$ values are not equal.

\begin{figure}
\begin{center}
\includegraphics[scale=0.5]{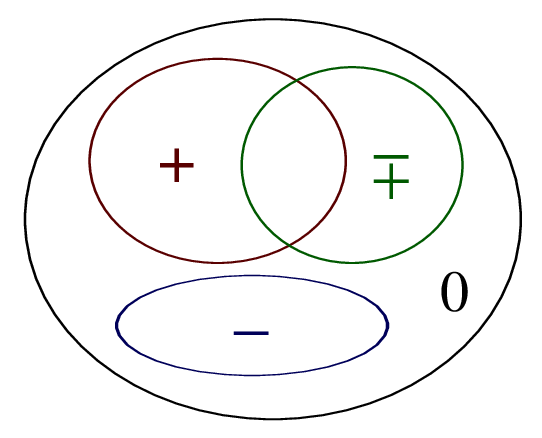}
\end{center}
\caption{\label{fig4}
Relationships between the sets of optimal measurements
for the four signs: $+$, $-$, $0$, and $\mp$.}
\end{figure}%
If two information--disturbance pairs have different signs,
their optimal measurements have different forms;
however, they partially overlap.
Figure~\ref{fig4} shows the relationships between
the sets of optimal measurements
for the four signs~\cite{Terash16}.
In this figure,
the identity operation and rank-$1$ projective measurement
are excluded because they are optimal in any case.
Table~\ref{tbl1} and Fig.~\ref{fig4} show that
the optimal measurements for $I$--$D_R$
are always non-optimal for $I$--$(1-R)$.
Although the optimal measurements for $G$--$(1-R)$
are not necessarily optimal for $G$--$(1-F)$~\cite{CheLee12,LRHLK14},
the optimal measurements for $I$--$D_R$ are
always optimal for $I$--$D_F$.

\section{\label{sec:summary}Summary}
Herein, we introduced additive measures for
information and disturbance in quantum measurements.
Three measures were defined from
the estimation fidelity, operation fidelity,
and physical reversibility using a logarithm.
By incorporating entropy reduction,
which is inherently additive,
four additive information--disturbance pairs were considered.
These pairs were derived
from two information and two disturbance measures.
For each information--disturbance pair,
we identified a physically allowed region
on the information--disturbance plane.
The region corresponding to a single measurement outcome
was obtained by
plotting all physically possible measurement operators
on the plane.
The convex hull of this region represents
the average values over the outcomes.
The lower boundary of this average region defines
an inequality that expresses the tradeoff relationship
between information and disturbance.

Furthermore, we found the optimal measurements
for each information--disturbance pair.
They correspond to the points on
the lower boundary for the average values
and are determined by the curvature of the lower boundary
of a single outcome.
As the lower boundary is deformed by the logarithm,
the optimal measurements were changed under the additivity.
In general,
information--disturbance pairs
can be classified according to the sign of the curvature.
For example, if two information--disturbance pairs have the same sign,
their optimal measurements take the same form.
Four cases---$+$, $-$, $0$, and $\mp$---were
analyzed for the additive and original measures.
The relationships between the sets of optimal measurements
for these four curvature signs were illustrated
in Fig.~\ref{fig4}.

The methods presented in this paper can be applied to
other multiplicative measures,
provided they can be defined for a single measurement outcome.
In future, exploring information--disturbance pairs
with signs beyond the four cases discussed (e.g., $\pm$)
or with lower boundaries that exhibit two or more inflection points
would be interesting.
Moreover,
the classification of information--disturbance pairs
can be extended to triplewise tradeoffs
using two signs characterizing
the curvature of a two-dimensional boundary.


\end{document}